\documentclass[11pt,a4paper]{article}
    \usepackage{jheppub}

\usepackage{amsmath, amsfonts, amsthm, amssymb, graphicx}
\usepackage{color}
\usepackage[utf8]{inputenc}
\usepackage{amsmath}
\usepackage{lipsum}
\usepackage{xcolor}
\usepackage{centernot}

\usepackage[normalem]{ulem}
\usepackage{feynmp}
\def\be{\begin{equation}}
\def\ee{\end{equation}}
\def\ba{\begin{eqnarray}}
\def\ea{\end{eqnarray}}
\newcommand{\mpl}{M_{P}}
\newcommand{\order}{{\cal O}}
\newcommand{\M}{{\cal M}}
\newcommand{\D}{{\cal D}}
\newcommand{\Q}{{\cal Q}}

\newcommand{\F}{{\cal F}}
\newcommand{\del}{\partial}
\def\d{{\rm d}}

\newcommand\myeq{\mathrel{\overset{\makebox[0pt]{\mbox{\normalfont\tiny\sffamily def}}}{=}}}

\newcommand{\citeseq}{\cite{KP1,KP2,KP3,KPSZ, KPS, KP4,etude,monseq,irat,omnia}}
\newcommand{\citecc}{\cite{zeldovich,wein,pol,Martin,cliff,me}}
\newcommand{\citemon}{\cite{gia1,gia2, KS,KSq,KLS,KLuv,Marchesano,Hebecker}}

\allowdisplaybreaks

\title{Quantum corrections to vacuum energy sequestering (with monodromy)}

\author[a]{Basem Kamal El-Menoufi,}
\emailAdd{b.elmenoufi@sussex.ac.uk}
\affiliation[a]{Department of Physics $\&$ Astronomy, 
University of Sussex, Falmer, Brighton, BN1 9QH, UK}
\author[b]{Silvia Nagy,}
\emailAdd{silvia.nagy@nottingham.ac.uk}
\affiliation[b]{School of Physics and Astronomy, 
University of Nottingham, Nottingham NG7 2RD, UK} 
\author[c]{Florian Niedermann,}
\emailAdd{niedermann@cp3.sdu.dk}
\affiliation[c]{CP$^3$-Origins, Center for Cosmology and Particle Physics Phenomenology, University of Southern Denmark, Campusvej 55, 5230 Odense M, Denmark} 
\author[b]{Antonio Padilla} 
\emailAdd{antonio.padilla@nottingham.ac.uk}

\date{\today}

\abstract{Field theory models of axion monodromy have been shown to exhibit vacuum energy sequestering as an emergent phenomenon for cancelling radiative corrections to the cosmological constant. We study one loop corrections to this class of models coming from virtual axions using a heat kernel expansion. We find that the structure of the original sequestering proposals is no longer preserved at low energies. Nevertheless, the cancellation of radiative corrections to the cosmological constant remains robust, even with the new structures required by quantum corrections.}

\begin{document}
\maketitle

\section{Introduction}
We begin with the cosmological constant problem \citecc. In quantum field theory,  the energy of the vacuum receives radiative corrections, generically scaling like the fourth power of the heaviest particle in the effective description. When  these fields are coupled to classical General Relativity, this vacuum energy, in accordance with the Equivalence Principle, gravitates like a cosmological constant.  Its UV sensitivity now becomes problematic since a typical field theory  cut-off  lies many orders of magnitude above the scale of dark energy.  Indeed, for a field theory with a TeV  cut off,   we must introduce a counterterm whose finite part cancels off  the finite part of the vacuum energy to one part in $10^{60}$ in order for the observed cosmological constant to be compatible with observational constraints. Such a tuning is, of course, radiatively unstable, and represents the worst failure of naturalness known to  Physics \cite{Giudice,Giudice2}.

A mechanism for tackling the cosmological constant problem has been developed in recent years, known as vacuum energy sequestering \citeseq.  The generic idea builds upon the fact that the cosmological constant is special because  it corresponds to an infinite wavelength source of energy and momentum. To modify its impact, it is sufficient to introduce  new rigid degrees of freedom, constant in both space and time, whose global dynamics force the desired cancellation of   radiative corrections.  In this sense, the Equivalence Principle is violated globally but not locally, the result of which  is an effective theory, locally equivalent to General Relativity, sourced by a radiatively stable cosmological constant.  The precise value of the cosmological constant is not predicted within the effective theory, but is set empirically. In this sense its status is similar to that of fermion masses in effective field theory: they are radiatively stable thanks to (approximate) chiral symmetry but their values are given by a measurement \cite{thooft}. The cancellation mechanism in sequestering models can also be understood in terms of so-called decapitation \cite{decap,selftune}.

 It is natural to ask whether or not the sequestering mechanism can emerge in a low energy effective theory arising from string compactifications.  To this end, it was very recently shown \cite{monseq,irat} how sequestering models can emerge from a pair of (deformed) field theory monodromies \citemon .  In field theory monodromy, there is a bilinear mixing between a scalar and a four-form field strength. The dynamics is that of a very massive scalar field and the  dual of the four form. The latter is constant in spacetime but can  play the role of a rigid degree of freedom familiar to sequestering.   Field theory monodromy set-ups  have been proposed as the low energy description of monodromy inflation in string theory \cite{mon1,mon2}, so we can be optimistic about finding a realisation of sequestering in a fundamental theory.
 
To gain further insight into what could emerge from string theory, we adopt a bottom up approach and consider  quantum corrections to monodromy models with an emergent sequestering mechanism.  In particular, we integrate out fluctuations in the two axion fields at one-loop {using heat kernel techniques to expand the effective action in powers of {\em derivatives}}. We find that the original constructions are not closed under renormalisation, even to a fixed order in the expansion. To carry out the renormalization program, we henceforth modify the original theory by including all necessary operators. With the renormalized theory in hand, we ask whether or not the sequestering mechanism is still operative. Remarkably, we find that under very reasonable assumptions on the parameters of the theory, the sequestering mechanism survives.

The rest of this paper is organised as follows. In section \ref{sec:review}, we review the emergence of the sequestering mechanism in the original deformations of a pair of field theory models of axion monodromy.  We generalise the results of \cite{monseq,irat}, slightly, allowing both axions to fluctuate and including different versions of low energy sequestering. In section \ref{sec:effaction}, we compute the quantum corrections to these theories due to virtual axions, at one loop,  using a heat kernel expansion.  We also elucidate the role of the global degrees of freedom at the level of the path integral. In the end we find corrections to the original models of \cite{monseq,irat}. In section \ref{sec:robust}, we explicitly demonstrate  how these corrections still allow for an emergent cancellation of radiative corrections to vacuum energy, via a generalisation of the sequestering mechanism.  We conclude in section \ref{sec:conc}.
 
 \section{Vacuum energy sequestering and monodromy} \label{sec:review}

 In \cite{monseq, irat}, it was shown how the sequestering mechanism could emerge from a pair of (deformed) monodromy constructions in field theory. Generalising slightly, these models could be described by the following action,
 \begin{eqnarray} \label{action}
S[A_i, \sigma_i, g, \Psi] =&&  \frac{M_P^2}{2}  \int d^{\text{4}}x \sqrt{g}  R+S_m [g_{\mu\nu}, \Psi]  \nonumber  \\
&&+ \sum_{i=1}^2\int d^{\text{4}}x \sqrt{g} \left[ -\frac{1}{2}  (\nabla\sigma_i )^2 + \frac{\mu_i}{4!}  \sigma_i \frac{\epsilon^{\mu\nu\alpha\beta}}{\sqrt{g}} F^i_{\mu\nu\alpha\beta}   - \frac{1}{2(4!)} F^i_{\mu\nu\alpha\beta} F_i^{\mu\nu\alpha\beta}   \right] \nonumber \\
&&+ \int d^{\text{4}}x \sqrt{g} \left[ \frac{M_P^{4-[O]}}{2} \mathcal{F}_O \left(\frac{\sigma_1}{M_P}\right) O[g] - \lambda^\text{4} \mathcal{F}_\lambda\left(\frac{\bar m \sigma_2}{\lambda^2}\right) \right] \,.
\end{eqnarray}
The first line is just the Einstein-Hilbert action with  matter fields, $\Psi$, minimally coupled to the metric, $g_{\mu\nu}$. The second line contains a pair of field theory monodromy constructions \citemon, with axions $\sigma_i$ and four-form field strengths, $F^i_{\mu\nu\alpha\beta}=4 \del_{[\mu} A^i_{\nu\alpha\beta]}$ whose bilinear mixing is controlled by the mass scale $\mu_i$. The final line contains the deformations that break the (discrete) shift symmetry for the axions,  with independent mass scales $\bar{m}$ and $\lambda$. Note that in the interests of brevity we have neglected to include boundary terms dependent on $g_{\mu\nu}$ and $\sigma_i$, required in order to have a well defined variational principle. Their precise form is non-trivial and specified in  \cite{KPS, omnia}, but not required for our analysis.

If we identify $\sigma_2$ with the inflaton, as in \cite{monseq}, one can generate the potential $\lambda^\text{4} \mathcal{F}_\lambda\left(\frac{\bar m \sigma_2}{\lambda^2}\right)$ from loops of matter coupled to the inflaton\footnote{Such couplings can be motivated by reheating \cite{monseq}.  For example,  suppose there is a coupling $\eta \sigma_2^2 h^2$ to some massive scalar $h$ of mass $m_h$ (possibly the Higgs), then loops of $h$ will generate a term of the form $\bar m^2 \sigma_2^2$ where  $\bar m \sim  \eta m_h $. For an effective theory cut-off at some scale ${\cal M} \sim 4\pi \lambda$, we can motivate the higher order operators using naive dimensional analysis \cite{nda1,nda2}, treating the mass, $\bar m$, as a spurion \cite{monseq}. }. The other deformation is assumed to be of gravitational origin, consistent with a breaking of the  shift symmetry induced by  gravitational instantons \cite{dine,wgc}.  Its precise form depends on the model of sequestering we wish to emerge at low energies. For earlier models of sequestering (see e.g.\ \cite{KPSZ}) we take  $O[g]=R$. For the more sophisticated later model \cite{KP4}, we take $O[g] =R_{GB}\myeq R_{\mu\nu\alpha\beta}R^{\mu\nu\alpha\beta}-4R_{\mu\nu}R^{\mu\nu}+R^2$ to be the Gauss-Bonnet combination. This model was recently dubbed {\it Omnia Sequestra} \cite{omnia} on account of its ability to sequester radiative corrections to vacuum energy from both matter {\it and} graviton loops. Note that our convention for the metric is $(-+++)$, the Riemann tensor is $R^\mu_{~\nu\alpha\beta} = \partial_\beta \Gamma^{\mu}_{\nu\alpha} - \cdots$ {and the Ricci tensor is $R_{\mu\nu} = R^\lambda_{~\mu\nu\lambda}$}.  We also denote the mass dimension of the operator $O$ by $[O]$.  
 
 To see how sequestering emerges at low energies, it is convenient to switch to a dual description by  integrating out the four-form field strengths and writing the action in terms of their magnetic duals. To do this, we add  a   pair of continuous Lagrange multipliers, $Q_i(x)$,  via a term  $\int d^4 x \frac{1}{4!} Q_i \epsilon^{\mu\nu \alpha \beta} (F^i_{\mu\nu\alpha\beta}-4 \del_{[\mu} A^i_{\nu\alpha\beta]})$, and include functional variation over both $Q_i$ and $F^i_{\mu\nu\alpha\beta}$. The four-forms now enter algebraically, and at quadratic order, so they can be integrated out exactly, yielding
  \begin{eqnarray} \label{action2}
S[A_i, \sigma_i, Q_i, g, \Psi]=&& \frac{M_P^2}{2}   \int d^{\text{4}}x \sqrt{g} R+S_m [g_{\mu\nu}, \Psi] \nonumber  \\
&&+ \sum_{i=1}^2\int d^{\text{4}}x \sqrt{g} \left[-\frac{1}{2}  (\nabla\sigma_i )^2 -\frac{1}{2} (\mu_i \sigma_i+Q_i)^2-\frac{1}{3!} Q_i \frac{\epsilon^{\mu\nu\alpha\beta}}{\sqrt{g}} \del_{[\mu} A^i_{\nu\alpha\beta]}   \right] \nonumber \\
&&+ \int d^{\text{4}}x \sqrt{g} \left[ \frac{M_P^{4-[O]}}{2} \mathcal{F}_O \left(\frac{\sigma_1}{M_P}\right) O[g] - \lambda^\text{4} \mathcal{F}_\lambda\left( \frac{\bar m \sigma_2}{\lambda^2}\right) \right] \, .
\end{eqnarray}
 The $Q_i$ are constrained to be  constant on-shell by the three-form variation. Furthermore, their expectation values are quantised in units of the membrane charge, $q_i$,   as in $\langle Q_i  \rangle=2\pi N q_i$ \cite{PS,BP}.  The second line of the action is manifestly invariant  under a simultaneous shift in $\sigma_i$ and $Q_i$. Non-perturbative effects break this down to a discrete symmetry \cite{callan}, consistent with the quantisation condition on the $ \langle Q_i  \rangle$, 
 \be
 \sigma_i \to \sigma_i+2\pi f_i, \qquad Q_i \to Q_i-2\pi q_i \,,
 \ee
 where the periodicity of the axions is given by $f_i=q_i/\mu_i$.  We now see how the axion, $\sigma_i$ has gained a mass, $\mu_i$,  from its bilinear mixing with the four-form field strength.  To connect directly to sequestering models, we {re-define $A_i \to A_i / \mu_i$ and} take the  limit where $\mu_i \to \infty$, $Q_i \to \infty$ holding $s_i=-Q_i/\mu_i$ fixed. In this limit  the axion settles  into its minimum at $s_i$ so that the action now reads
 \begin{eqnarray} 
S[A_i, \sigma_i, g, \Psi] =&&  \frac{M_P^2}{2}  \int d^{\text{4}}x \sqrt{g} R+S_m [g_{\mu\nu}, \Psi]  \nonumber  \\
&&+ \sum_{i=1}^2\int d^{\text{4}}x \sqrt{g} \left[ \frac{1}{3!}  s_i \frac{\epsilon^{\mu\nu\alpha\beta}}{\sqrt{g}} \del_{[\mu} A^i_{\nu\alpha\beta]}   \right] \nonumber \\
&&+ \int d^{\text{4}}x \sqrt{g} \left[ \frac{M_P^{4-[O]}}{2} \mathcal{F}_O \left(\frac{s_1}{M_P}\right) O[g] - \lambda^\text{4} \mathcal{F}_\lambda\left(\frac{\bar m s_2}{\lambda^2}\right) \right] \,. \label{limit}
\end{eqnarray}
 This is precisely the form of the  actions presented in \cite{KPSZ, KP4}, suggesting that the low energy dynamics of our deformed monodromies \eqref{action} is equivalent to a sequestering theory.  Furthermore,  given that the cancellation of vacuum energy occurs at infinite wavelength in sequestering, we might expect it to still occur even in the full monodromy theory \eqref{action}. We shall now show that this is indeed the case.
 
 We start with the action \eqref{action2} and go to the global limit by integrating out the three-forms (see appendix \ref{app:forms}).  This yields
   \begin{eqnarray} \label{action3}
S[ \sigma_i,  g, \Psi ; Q_i) =&& \frac{M_P^2}{2}   \int d^{\text{4}}x \sqrt{g} R+S_m [g_{\mu\nu}, \Psi] \nonumber  \\
&&+ \sum_{i=1}^2-Q_i C_i +\int d^{\text{4}}x \sqrt{g} \left[ -\frac{1}{2}  (\nabla\sigma_i )^2 -\frac{1}{2} (\mu_i \sigma_i+Q_i)^2\right]\nonumber \\
&&+ \int d^{\text{4}}x \sqrt{g} \left[ \frac{M_P^{4-[O]}}{2} \mathcal{F}_O \left(\frac{\sigma_1}{M_P}\right) O[g] - \lambda^\text{4} \mathcal{F}_\lambda\left(\frac{\bar m \sigma_2}{\lambda^2}\right) \right] \,,
\end{eqnarray}
 where the $Q_i$ are assumed to be global Lagrange multipliers constant in spacetime and $C_i \myeq \int_\M dA_i=\int_{\del \M} A_i$ is the integral of the three-form flux across the  boundary, $\del \M$, of the spacetime, $\M$. Our notation reflects the fact that this action is a functional of the fields $\sigma_i,  g, \Psi$, but a function of the global variables $Q_i$. Variation with respect to  the axions and the metric  yields the following local equations
\begin{subequations} 
 \begin{eqnarray}
 \frac{M_P^{3-[O]}}{2} \mathcal{F}_O' \left(\frac{\sigma_1}{M_P}\right) O[g] &=& -\square \sigma_1+{\mu_1} \left( \mu_1\sigma_1+Q_1 \right)\label{sc1} \,,\\
 -\bar m \lambda^2  \mathcal{F}_\lambda'\left(\frac{\bar m \sigma_2}{\lambda^2 }\right) &=&  -\square\sigma_2+{\mu_2} \left(\mu_2\sigma_2+Q_2\right) \label{sc2} \,,\\
 \mpl^2 G_{\mu\nu}+M_P^{4-[O]} E^O_{\mu\nu}(\mathcal{F}_O) &=& -\lambda^4 \mathcal{F}_\lambda g_{\mu\nu} +T_{\mu\nu} \nonumber \\
 &&  +\sum_{i=1}^2 2 \nabla_\mu \sigma_i \nabla_\nu \sigma_i-g_{\mu\nu} \left[(\nabla\sigma_i )^2+\frac{1}{2} (\mu_i \sigma_i+Q_i)^2\right]  \label{g} \,,
 \end{eqnarray}
 \end{subequations}
 where $T_{\mu\nu}=-\frac{2}{\sqrt{g}} \frac{\delta S_m}{\delta g^{\mu\nu}}$ is the energy-momentum tensor in the matter sector  and 
  \begin{eqnarray}
 E^O_{\mu\nu}(\mathcal{F}_O) &&= \frac{1}{\sqrt{g}} \frac{\delta}{\delta g^{\mu\nu}}  \int d^{\text{4}}x \sqrt{g} \mathcal{F}_O \left(\frac{\sigma_1}{M_P}\right)  O[g]  \nonumber \\
 &&=\begin{cases} \left(-\nabla_{\mu} \nabla_{\nu} +g_{\mu\nu} \square +G_{\mu\nu} \right)\mathcal{F}_O & \text{for~} O[g]=R \\
 -4P_{\mu\alpha\nu\beta} \nabla^\alpha \nabla^\beta \mathcal{F}_O  &  \text{for~}  O[g]=R_{GB}  \end{cases}
  \end{eqnarray}
  is the variation of the gravity deformation (see e.g.  \cite{fab4}).  Here $P_{\mu\alpha\nu\beta} $ is the double dual of the Riemann tensor.  The energy-momentum tensor for the axions appears in the last terms of \eqref{g}.  Variation with respect to the global Lagrange multipliers yields the following global constraints,
  \be \label{C}
  C_i =-\int d^4 x \sqrt{g} (\mu_i\sigma_i+Q_i) \,.
 \ee
Since  $C_i \myeq \int_\M dA_i=\int_{\del \M} A_i$, if we were to redefine the three-forms  $A_i \to (A_i-\star d \sigma_i)/{\mu_i}$, where $\star$ is the Hodge dual, then $C_i \to (C_i-\int d^4 x \sqrt{g} \square \sigma_i)/{\mu_i}$,  and we can rewrite these global constraints as
 \be \label{Cmod}
 C_i=-\int d^4 x \sqrt{g} \left[-\square \sigma_i+{\mu_i} \left(\mu_i\sigma_i+Q_i\right)\right]\,.
 \ee 
 Integrating the scalar equations of motion  \eqref{sc1} and \eqref{sc2} over spacetime and taking the ratio yields a global  geometric constant which can be written as
 \be
 \langle O \rangle_w=-2\bar m \lambda^2 \mpl^{[O]-3} \left\langle\frac{ \mathcal{F}_\lambda'}{ \mathcal{F}_O' } \right \rangle_w \frac{C_1}{C_2}\,, \label{globconstraint}
 \ee
 where  we have introduced the weighted spacetime average $ \langle Q \rangle_w=\frac{\int d^4 x \sqrt{-g} w Q}{\int d^4 x \sqrt{-g} w} $, and we are taking $w=\mathcal{F}_O' $. This constraint is crucial. As long as the operator $O$ depends on the scale dependent part of the curvature, the constraint \eqref{globconstraint} fixes the large wavelength mode of the scalar curvature in a radiatively stable way \cite{KP4,etude}.  At the level of the gravity equation \eqref{g}, we identify the long wavelength mode of $\lambda^4 \mathcal{F}_\lambda+\sum_i \frac{1}{2} (\mu_i \sigma_i+Q_i)^2$ with the cosmological constant counterterm, whose value is dynamically fixed by the constraint \eqref{globconstraint}. Indeed, we can eliminate $\langle \lambda^4 \mathcal{F}_\lambda +\sum_i \frac{1}{2} (\mu_i \sigma_i+Q_i)^2 \rangle_w$ from \eqref{g} by taking the trace and (weighted) spacetime average, allowing us to  rewrite the gravity equation as
 \begin{multline}
  M_\text{eff}^2 G_{\mu\nu}+M_P^{4-[O]} E^O_{\mu\nu}(\delta \mathcal{F}_O) =T_{\mu\nu}-\frac14 \langle T\rangle_w g_{\mu\nu} -\delta \left[\lambda^4 \mathcal{F}_\lambda+ \sum_{i=1}^2 \frac{1}{2} (\mu_i \sigma_i+Q_i)^2\right] g_{\mu\nu} \\+\sum_{i=1}^2  \left[2 \nabla_\mu \sigma_i \nabla_\nu \sigma_i-g_{\mu\nu} (\nabla\sigma_i )^2  \right]- \Lambda_\text{global} g_{\mu\nu} \,, \label{effg}
 \end{multline}
where $\delta \chi= \chi-\langle  \chi  \rangle_w$ is the {\it localised} fluctuation in a generic field $\chi$  and  
 \be
M_\text{eff}^2 =\begin{cases} \mpl^2(1+\langle \mathcal{F}_O\rangle_w) & \text{for~} O[g]=R \\  \mpl^2   &  \text{for~}  O[g]=R_{GB}  \end{cases} 
\ee
is the effective Planck mass. Notice that we have split the effective potential into two parts: the localised part, corresponding to  $\delta \left[ \lambda^4 \mathcal{F}_\lambda +\ldots\right]$, and the global, infinite wavelength part, $\Lambda_\text{global}$.  The latter is given by 
\be
\Lambda_\text{global}=\Lambda_*-\frac14\left[M_P^{4-[O]}  \langle  E^O(\delta \mathcal{F}_O) \rangle_w+2\sum_{i=1}^2 \left \langle( \nabla {\sigma_i} )^2 \right\rangle_w \right] \,, \label{Lglob}
\ee
where  $E^O=g^{\mu\nu} E^O_{\mu\nu}$ and 
\be \label{Lstar}
\Lambda_*=\begin{cases} \frac14 M_\text{eff}^2  \langle O \rangle_w  & \text{for~} O[g]=R \\
\pm  \frac14 \mpl^2 \sqrt{6  \langle O \rangle_w -  6 \left\langle W_{\mu\nu\alpha\beta}^2  -2\left(R_{\mu\nu}-\frac14 R g_{\mu\nu}\right)^2 + {\frac{1}{6}}(\delta R)^2 \right \rangle_w}  &  \text{for~}  O[g]=R_{GB} 
\end{cases}
\ee 
Derivation of the  last relation uses the fact we can write the Gauss-Bonnet term as $R_{GB}=W_{\mu\nu\alpha\beta}^2  -2\left(R_{\mu\nu}-\frac14 R g_{\mu\nu}\right)^2 +\frac16 R^2$, where $W_{\mu\nu\alpha\beta}$ is the Weyl tensor \cite{KP4}. Let us first consider radiative corrections from the matter sector and their backreaction on to the geometry. To see this cleanly, it is convenient  to go to the global limit and consider maximally symmetric configurations of constant curvature, $R_{\mu\nu\alpha\beta}=\frac{\bar{R}}{12} (g_{\mu\alpha} g_{\nu\beta}- g_{\mu\beta} g_{\nu\alpha})$,  constant scalars, $\sigma_i=\bar \sigma_i$, and a vacuum energy source, $T_{\mu\nu}=-V_\text{vac} g_{\mu\nu}$. Plugging this ansatz into the effective gravity equation \eqref{effg} yields 
\be \label{barR}
\bar R=\begin{cases}   \langle O \rangle_w  & \text{for~} O[g]=R \\
\pm   \sqrt{6  \langle O \rangle_w}  &  \text{for~}  O[g]=R_{GB} 
\end{cases}
\ee
It follows that the curvature is stable against radiative corrections from the matter sector provided $\langle O \rangle_w$ is radiatively stable, in both cases. However, $\langle O \rangle_w$ is constrained by the fluxes according to equation  \eqref{globconstraint}, so the question is  now whether or not the right hand side of  \eqref{globconstraint} is radiatively stable. The fluxes $C_i$ are set as as infra-red boundary condition and have no ultra-violet sensitivity. In contrast, radiative corrections {\it will} affect the values for the scalars, $\bar \sigma_i$, and this could contaminate $\langle O \rangle_w$   through the averaged prefactor, going  as $ \frac{ \mathcal{F}_\lambda' \left(\frac{\bar m \bar\sigma_2}{\lambda^2}\right)}{ \mathcal{F}_O'\left(\frac{\bar \sigma_1}{M_P}\right)}
$ in our global limit. The radiative stability of this prefactor can be inferred from corrections to  the  $O[g]$-coupling, $ \frac{M_P^{4-[O]}}{2} \mathcal{F}_O \left(\frac{\bar \sigma_1}{M_P}\right)$ and  to  the cosmological constant ``counterterm", $ \lambda^4 \mathcal{F}_\lambda \left(\frac{\bar m \bar\sigma_2}{\lambda^2}\right)$. For a theory cut-off at some ultra-violet scale, $\mathcal{M}$, these go as \cite{Demers}
\begin{eqnarray}
\frac{M_P^{4-[O]}}{2} \mathcal{F}_O \left(\frac{\bar \sigma_1}{M_P}\right)  &&\to  \frac{M_P^{4-[O]}}{2} \mathcal{F}_O \left(\frac{\bar \sigma_1}{M_P}\right) +\order\left(\frac{\mathcal{M}^2}{{4 \pi} }\right)^{2-\frac{[O]}{2}}\log\left(\frac{\mathcal{M}}{m}\right) \\
  \lambda^4 \mathcal{F}_\lambda \left(\frac{\bar\sigma_2}{\lambda}\right) &&\to  \lambda^4 \mathcal{F}_\lambda \left(\frac{\bar\sigma_2}{\lambda}\right)+\order \left(\frac{\mathcal{M}^4}{16 \pi^2} \right)\log\left(\frac{\mathcal{M}}{m}\right)
\end{eqnarray}
where $m$ is a typical mass scale in the effective field theory governing the matter sector. Since $[O] \leq 4$,  radiative stability of $ \mathcal{F}_O $ requires  it to go at least as $\order(1) \left(\frac{\mathcal{M}^2}{4\pi \mpl^2}\right)^{2-\frac{[O]}{2}}$, which is expected to be the case  provided $\mathcal{M} \lesssim \mpl$. Similarly,  radiative stability of $\mathcal{F}_\lambda$ requires it to go as  $\order(1) \frac{\mathcal{M}^4}{16 \pi^2 \lambda^4 }$. This is also expected to be the case precisely because $ \lambda^4 \mathcal{F}_\lambda$  plays the role of the counter-term cancelling off radiative corrections to vacuum energy going as $\order \left(\frac{\mathcal{M}^4}{16 \pi^2}\right)$. For generic smooth functions $\mathcal{F}_O$ and $\mathcal{F}_\lambda$, this stability is transferred to the prefactor $ \frac{ \mathcal{F}_\lambda' \left( \frac{\bar  m\bar \sigma_2}{\lambda^2}\right)}{ \mathcal{F}_O'\left(\frac{\bar \sigma_1}{M_P}\right)}$. We therefore see that  $\langle O \rangle_w $ is radiatively stable against matter loops, and by association the same must be true for the long wavelength mode of the spacetime  curvature, $\bar R$. 

Although we have focussed on the global limit, this is just for convenience of presentation. The local fluctuations are irrelevant when considering the gravitational effects of vacuum energy since they cannot be sourced by it. That is not to say the local fluctuations do not contribute to the geometry on large scales. They can, through their (weighted) spacetime averages entering in \eqref{Lglob}. However, these contributions are immune from the radiative instabilities contained in the vacuum energy. For the monodromy set-up studied here, the local fluctuations can, of course, backreact on to the geometry on shorter scales.  For the case of $\sigma_2$, such fluctuations can even give rise to inflation, as suggested in \cite{monseq}.  Fluctuations in $\sigma_1$ are potentially more exotic since they represent a departure from local General Relativity \cite{mg}.  As long as we make the mass of the fluctuations, $\mu_1$, sufficiently large, beyond the scale of inflation,  there will be no conflict with observation. 

Let us now turn our attention to graviton loops. This is where our two models go their separate ways.  For $O[g]=R$, as in earlier sequestering proposals \cite{KPSZ}, graviton loops can and will induce a UV sensitive potential for $\sigma_1$. This is problematic because it will source equation \eqref{sc1} and contaminate the constraint \eqref{globconstraint} \cite{KP4}. In contrast, when   $O[g]=R_{GB}$, there is an approximate shift symmetry $\mathcal{F}_O \to \mathcal{F}_O+c$, becoming exact in the limit $\mu_1, Q_1 \to \infty,  \mu_1C_1 \to 0$, holding $Q_1/\mu_1$ fixed, as we can easily see by staring at the limiting action \eqref{limit}.  The shift symmetry is also exact even without taking  these limits, provided we take $\mathcal{F}_O$ to be linear. In any event, this shift symmetry, exact or otherwise, helps protect us from a perturbatively generated  UV sensitive potential for $\sigma_1$ and contamination of the constraint \eqref{globconstraint}.  Non-perturbative effects will break the symmetry down to discrete shifts, of course, although these corrections will be suppressed for large instanton action \cite{artur}.

The UV completion of sequestering via monodromy brings in another source of radiative corrections, from axion loops.  The main concern, here, is that the action \eqref{action} is not closed under quantum corrections, and that to fix this we spoil the cancellation mechanism that renders the observed cosmological constant radiatively stable. We will investigate this in detail in the next two sections.

\section{The quantum effective action} \label{sec:effaction}
To what extent does the sequestering mechanism survive radiative corrections from the gravitational sector?  A complete analysis requires us to consider the effects of quantum fluctuations of all fields, including gravitons, axions, 3-forms and, of course, the matter fields.  As a first step, here we will integrate out the fluctuations in the axions at one loop and compute the corresponding one-loop effective action for sequestering.  The metric will be treated  as a classical background field while quantum fluctuations in the matter sector will only be considered through their contributions to the vacuum energy.  We will neglect sources for the 3-forms allowing us to integrate them out exactly.

Our starting point will  be the action  \eqref{action}. The path integral is given by the functional integration over the dynamical fields
\be
Z[J_i]={\cal N} \int \ldots  \D\sigma_i   \D A_{\mu\nu\alpha} ^i ~e^{iS[A_i, \sigma_i, \ldots]+i \int d^4 x \sqrt{-g} J_i  \sigma_i } \,,
\ee
where ${\cal N}$  is a constant normalisation factor. We have focussed on the integration over the  axions and  the 3-forms, although as stated above,  here we only consider sources for the axions.  The action $S[A_i, \sigma_i, \ldots]$ is given by \eqref{action}. 
It also includes  boundary terms  in order to have a well defined variational principle. As explained in  \cite{KPS, omnia}, for the sequestering cancellation to occur we need to retain the global variations in the (pseudo) scalars. This means we must impose Neumann, rather than Dirichlet, boundary conditions on them, which in  turn requires us to adjust the boundary condition on the metric\footnote{For $O[g]=R$ one must impose Dirichlet boundary conditions on the Einstein frame metric, which is  related Jordan frame metric by a conformal transformation \cite{KPS}. A similar, but more complicated boundary condition is also required for $O[g]=R_{GB}$ \cite{omnia}.}. {In what follows, we assume the action to be supplemented with boundary terms necessary for the heat kernel method to be well defined given our choice of boundary conditions.}

To proceed, we pass to the dual description described in the previous section. To do so, recall that we promote $F^i_{\mu\nu\alpha\beta}$ to a field,  introduce a Lagrange multiplier, $Q_i$,  to impose  the constraint, $\frac{1}{4!}  \epsilon^{\mu\nu \alpha \beta} (F^i_{\mu\nu\alpha\beta}-4 \del_{[\mu} A^i_{\nu\alpha\beta]})=0$, and include functional variation over both $Q_i$ and $F^i_{\mu\nu\alpha\beta}$. Since the $F^i_{\mu\nu\alpha\beta}$ now enters the action up to quadratic order we are able to integrate it out exactly, giving the action  \eqref{action2} and an overall constant normalisation in the path integral, which we specify in the appendix \ref{app:forms}. The  result is 
\be
Z[J_i]={\cal N}' \int \ldots  \D\sigma_i  \D Q_i \D A_{\mu\nu\alpha} ^i ~e^{iS[A_i, \sigma_i, Q_i \ldots]+i \int d^4 x \sqrt{-g} J_i \sigma_i  } \,,
\ee
where ${\cal N}'$  is,  again, a constant normalisation factor  and $S[A_i, \sigma_i, Q_i \ldots]$ is given by \eqref{action2}.  Now, as explained in detail in the appendix \ref{app:forms}, performing the 3-form integration yields a delta functional, forcing the gradient, $\del_\mu Q_i$, to vanish. This suppresses local, but not global,  off-shell fluctuations in the $Q_i$, reducing the path integration over fields, $Q_i(x)$, to an ordinary integration over the global variables, $Q_i$.  As a result, we now work with the action of the form \eqref{action3} and a path integral $Z[J_i]=\int   d Q_i Z_{Q_i} [J_i]$ with\footnote{The constant normalisation factor that arises from integrating out the 3-forms can be absorbed in $\mathcal{N}'$.}
\be
 Z_{Q_i} [J_i] ={\cal N}' \int \ldots  \D\sigma_i   ~e^{iS[\sigma_i,\ldots ; Q_i)+i \int d^4 x \sqrt{-g} J_i \sigma_i  } \,.
\ee
This structure is reminiscent of a gas of ``Universes" whose coupling constants are random variables to be summed over (see e. g. \cite{Cole,Banks}). Focussing on one such ``Universe" (with $Q_i$ fixed) we now integrate out the fluctuations in the axions   at one loop. To this end, we derive the  resulting quantum effective action using a  heat kernel expansion. Rather than discussing both sequestering models in turn as done in the previous section, we will study a generalised version of action \eqref{action3} which couples $\sigma_1$ to both $\mathcal{O}[g]=R$ and $\mathcal{O}[g]=R_{GB}$. If we want to make contact with the original models, we can project on the corresponding parameter subspace whenever deemed necessary. Anticipating the final result, in order to renormalize the theory we are also forced to generalize the action to include a coupling between $\sigma_2$ and curvature and thus we have instead
\begin{align} \label{action4}
\nonumber
S[ \sigma_i,  g, \Psi ; Q_i) =& \frac{M_P^2}{2} \int d^{\text{4}}x \sqrt{g}  Z R+S_m [g_{\mu\nu}, \Psi]   + {S_{\partial \mathcal{M}}[g_{\mu\nu}, \sigma_i] }\\\nonumber
&+ \sum_{i=1}^2-Q_i C_i +\int d^{\text{4}}x \sqrt{g} \left[ -\frac{1}{2}  (\nabla\sigma_i )^2 -\frac{1}{2} (\mu_i \sigma_i+Q_i)^2 \right]\\\nonumber 
&+ \int d^{\text{4}}x \sqrt{g} \bigg[ \frac{M_P^{2}}{2} \mathcal{F}_R\left(\frac{\sigma_1}{M_P}, \frac{\sigma_2}{M_P}\right) R +  \frac12 \mathcal{F}_{R_\text{GB}} \left(\frac{\sigma_1}{M_P}, \frac{\sigma_2}{M_P}\right) R_{\text{GB}} \\
&- \lambda^\text{4} \mathcal{F}_\lambda\left(\frac{\bar m \sigma_2}{\lambda^2}\right) \bigg]\,.
\end{align}
Here we treat $\mpl, \lambda, \bar m$ as {\it fixed scales} and capture the RG flow in terms of dimensionless couplings, such as $Z$, or coefficients in the functional definition of the $\mathcal{F}_R, \mathcal{F}_{\text{GB}}$ and $\F_\lambda$.  Note that both $\mathcal{F}_R$ and $\mathcal{F}_{\text{GB}}$ retain full dependence on both axions, except that we shall still assume the absence of any direct mixing {(see the discussion in section \ref{sec:conc})}, so that
\begin{subequations}
\label{FO}
\begin{align} 
\mathcal{F}_R \left(\frac{\sigma_1}{M_P}, \frac{\sigma_2}{M_P}\right) &=\mathcal{F}_{R,1} \left(\frac{\sigma_1}{M_P}\right) + \mathcal{F}_{R,2} \left(\frac{\sigma_2}{M_P}\right) \,,\\
\mathcal{F}_{R_\text{GB}} \left(\frac{\sigma_1}{M_P}, \frac{\sigma_2}{M_P}\right) &=\mathcal{F}_{R_\text{GB}, 1} \left(\frac{\sigma_1}{M_P}\right)+\mathcal{F}_{R_\text{GB}, 2} \left(\frac{\sigma_2}{M_P}\right)\,.
\end{align}
\end{subequations}
For a generic scalar field, $\sigma$,  in $d$ dimensional curved space, the effective action can be expressed as the trace of the heat kernel
\begin{align}\label{EAfund}
\Gamma_{\text{1-loop}}[g] = -\frac{i\hbar}{2} \int d^{\text{d}}x \int_0^\infty \frac{ds}{s} H(x,x;s) \,,
\end{align}
where the heat kernel is a solution to a heat-diffusion equation
\begin{align}
(\partial_s + \mathcal{D}_x) H(x,y;s) = 0, \quad H(x,y;0) = \delta^{(d)}(x,y)\,,
\end{align}
and as such is defined as a bi-tensor density of weight-$1/2$ and $\mathcal{D}$ is a second-order differential operator \text{whose eigenmodes are scalar densities of weight-$1/2$}. If the operator contains a constant mass term, for example $\mathcal{D} = \mathcal{O} + m^2$, we have
\begin{align}
H(x,y;s) = e^{-m^2 s} \tilde{H}(x,y;s), \quad (\partial_s + \mathcal{O}_x) \tilde{H}(x,y;s) = 0 \ \ .
\end{align}
For the axion in \eqref{action4}, the operator of interest takes the simple form
\begin{align}\label{opform}
\mathcal{O} =- g^{\mu\nu} \nabla_\mu \nabla_\nu + {\cal Q}(x) \,,
\end{align}
where $\Q(x)$ is a space-time dependent function of the metric and all background fields. In each case we have a candidate mass parameter given by the $\mu_i$. Formally, ultraviolet divergences in the effective action (\ref{EAfund}) are controlled by the small $s$ behaviour of $\tilde{H}(x,y;s)$, and we can use the asymptotic expansion, as $s \to 0$, of the heat kernel; see \cite{Vas2003} and references therein. Explicitly, the trace takes the form
\begin{align}\label{DWexp}
\tilde{H}(x,x;s) \sim \frac{i}{(4\pi s)^{d/2}} \sum_{k=0}^{\infty} s^k E_k(x) \ \ ,
\end{align}
where the $\sim$ sign reflects the fact that this is an asymptotic expansion.
The central problem then reduces to computing the coefficient functions $E_k$ for a given operator. Let us first note that
\begin{align}
\int_0^\infty \frac{ds}{s} e^{-m^2 s} \frac{s^k}{(4\pi s)^{d/2}} = \frac{1}{16\pi^2} \left(1 - \epsilon \ln \left(\frac{m^2}{4\pi}\right) \right) (m^2)^{2-k} \,\Gamma[-2+k+\epsilon] \,,
\end{align}
where we used $d=4-2\epsilon$. Now we see the general form of the effective action. The UV divergences arise from ($E_0,E_1,E_2$) in the expansion of Eq.~(\ref{DWexp}). The coefficients $E_3(x)$ and higher yield finite contributions to the effective action, which are suppressed by the field mass. The celebrated Seeley-Gilkey-de Witt coefficients read \cite{Vas2003}
\begin{align}
E_0(x) &= \sqrt{g}, \quad E_1(x) = \sqrt{g} \left(\frac{R}{6} - \Q \right) \\ 
E_2(x) &=  \sqrt{g} \left( \frac{1}{30} \Box R + \frac{1}{120} R^2 {+} \frac{1}{60} R_{\mu\nu}R^{\mu\nu} + \frac{1}{180} R_{GB} + \frac{1}{2} \Q^2 - \frac16 R \Q - \frac16 \Box \Q \right) \\\nonumber
E_3(x) &= \sqrt{g} \bigg( \{\mathcal{R}^3, \nabla \nabla \mathcal{R}^2, \Box \Box R \} + \frac{1}{360} \bigg( 18 \Q \Box R - 60 \Q \Box \Q + 6 \Box \Box \Q + 60 \Q^3 \\
&+ 30 \Q^2 R + 5 \Q R^2 - 2 \Q R_{\mu\nu} R^{\mu\nu} + 2 \Q   R_{\mu\nu\alpha\beta} R^{\mu\nu\alpha\beta} \bigg) \bigg) \label{E3} \ \ ,
\end{align} 
where $\{\mathcal{R}^3, \nabla \nabla \mathcal{R}^2, \Box \Box R \}$ denotes the set of all purely geometric invariants containing six derivatives. 

Given the form of the action \eqref{action4}, for each axion,  $\sigma_i$,   we obtain the differential operators 
\be \label{operator}
\D_i= -g^{\mu\nu} \nabla_\mu \nabla_\nu+\Q_i +\mu_i^2 \,,
\ee
where
\begin{eqnarray}
\Q_1 &=& -\frac{1}{2}  \mathcal{F}_{R, 1}'' \left(\frac{\sigma^0_1}{M_P}\right) R -\frac{1}{2 M_P^2}  \mathcal{F}_{R_{\text{GB}}, 1}'' \left(\frac{\sigma^0_1}{M_P}\right) R_{\text{GB}}  \\
\Q_2 &=& -\frac{1}{2} \mathcal{F}_{R,2}'' \left( \frac{\sigma^0_2}{M_P}\right) R  -\frac{1}{2 M_P^2}  \mathcal{F}_{R_{\text{GB}}, 2}'' \left(\frac{\sigma^0_2}{M_P}\right) R_{\text{GB}}+ \bar m^\text{2} \mathcal{F}''_\lambda\left(\frac{\bar m \sigma^0_2}{\lambda^2}\right) \ \ .
\end{eqnarray}
where the superscript $0$ denotes background values, nevertheless, we will drop it from now on to simplify notation. {It is essential to notice that, regardless of the boundary conditions imposed on quantum fluctuations, the differential operator in Eq.~(\ref{operator}) is self-adjoint with respect to the natural inner product on the space of real functions \cite{Vas2003}. Moreover, any boundary terms resulting due to integration by parts can be made to vanish under the relevant boundary conditions.}
With the above ingredients, we re-write the effective action as follows
\begin{align} \label{effact}
\Gamma_{\text{1-loop}}[g] = \frac{\hbar}{32\pi^2} \sum_{i=1}^2 \int d^{\text{d}}x  \sum_{k=0}^\infty \left(1 - \epsilon \ln \left(\frac{\mu_i^2}{4\pi}\right) \right) (\mu_i^2)^{2-k} \,\Gamma[-2+k+\epsilon] E_k(x;\Q_i) \ \ .
\end{align}
In what follows,  we shall use the following replacement
\begin{align}
\int d^{\text{d}}x \to \mu^{2\epsilon} \int d^{4}x
\end{align}
to pass to 4 dimensions where $\mu$ is the scale of dimensional regularization. We shall also use the $\overline{\text{MS}}$ renormalization scheme and define
\begin{align}
\frac{1}{\bar{\epsilon}} := \frac{1}{\epsilon} - \gamma_E + \ln(4\pi) \ \ .
\end{align}
where $\gamma_E$ is the Euler-Mascheroni constant. As stated above, the $k=0, 1, 2$ contributions in \eqref{effact} are divergent and in general we obtain, 
\begin{multline} \label{effact1}
\Gamma_{\text{1-loop}}[g] = \frac{\hbar}{32\pi^2} \sum_{i=1}^2 \int d^{\text{4}}x \left\{
 \frac{\mu_i^4}{2} \left[\frac{1}{\bar{\epsilon}} + \frac32 - \ln \left(\frac{\mu_i^2}{\mu^2}\right)\right] - {\mu_i^2} \left[\frac{1}{\bar{\epsilon}} +1 - \ln \left(\frac{\mu_i^2}{\mu^2}\right)\right] E_1(x;\Q_i) \right. \\
 \left.\qquad  +\left[\frac{1}{\bar{\epsilon}}  - \ln \left(\frac{\mu_i^2}{\mu^2}\right)\right] E_2(x;\Q_i)
 +\sum_{k=3}^\infty (\mu_i^2)^{2-k} \,(k-3)! E_k(x;\Q_i) \right\}
\end{multline}
Now, we have to specify the exact forms of the functions $\mathcal{F}_{R,i}$, $\mathcal{F}_{\text{GB},i}$ and $\mathcal{F}_{\lambda}$. We elucidated in the previous section that these functions must not be linear, but otherwise arbitrary; accordingly,
\begin{subequations}
\label{ansatzFs}
\begin{align}
\mathcal{F}_\lambda(X)&= \alpha+\sum_{k\geq1} \beta_k X^k\,, \\
 \mathcal{F}_{R,i} (X)&= \sum_{k\geq1} c_k^{(i)}  X^k \,,\\
 \mathcal{F}_{R_{\text{GB}},i} (X)&= \sum_{k\geq1} d_k^{(i)}  X^k  \,.
\end{align}
\end{subequations}
{In the following we set $c_{k\geq 3}^{(i)}=0$ and $\beta_{k \geq 5} =0$.\footnote{{If we were to add these terms, we also would need to include couplings between $\sigma_i$ and $R^2$ to be able to renormalise the theory, which would somewhat complicate our further discussion.}} We will see that within our analysis this is a self-consistent choice as it is not spoiled by axion loops. Let us stress though that these operators (as well as those describing a mixing between both axion sectors or an axion coupling to higher curvature invariants) are expected to be produced by quantum gravity corrections and are therefore generically present in an effective theory of sequestering. Here, on the other hand, we are interested in a  first nontrivial check of the  model's stability under axion loops rather than a exhaustive analysis of all EFT operators. Correspondingly, we only discuss the simplified version of \eqref{ansatzFs} and speculate later about the relevance of the neglected operators (see section \ref{sec:conc} for a more detailed discussion of this point).}

We further amend the effective action to include terms quadratic in curvature, 
\begin{align} \label{R2}
\mathcal{S}_{R^2} = \int d^4x \sqrt{g} \left(u R^2 + v R_{\mu\nu} R^{\mu\nu}  \right) \,,
\end{align}  
where $u, v$ are the bare Wilson coefficients. We can now plug back in \eqref{effact1} and simply determine the renormalization constants which are given in appendix \ref{CTls1} by Eqs.\ \eqref{deltas}. The corresponding beta functions of the various couplings are listed in the appendix as well. We take as our initial conditions for the RGEs the values of the couplings at some scale $\mu_\star$. It is indeed possible to solve the coupled set of RGEs, nevertheless, we choose not to perform the explicit computation since it is not essential for our purposes. Hence, we finally arrive at the renormalised sequestering action correct up to one loop {(suppressing boundary terms)}
\begin{multline} \label{actionren}
S_\text{ren}[ \sigma_i,  g, \Psi ; Q_i) = \int d^{\text{4}}x \sqrt{g}   \left[ \frac{M_P^2}{2} Z^{\text{eff}}(\mu_\star)  R+u^{\text{eff}}(\mu_\star) R^2 + v^{\text{eff}}(\mu_\star) R_{\mu\nu} R^{\mu\nu}  
\right]+S_m [g_{\mu\nu}, \Psi]   \\
+ \int d^{\text{4}}x \sqrt{g} \left[ \frac{M_{P}^{2}}{2}\left( \mathcal{F}^{\text{eff}}_{R, 1} \left(\frac{\sigma_1}{M_{P}} ; \mu_\star\right)+\mathcal{F}^{\text{eff}}_{R, 2} \left(\frac{\sigma_2}{M_{P}} ; \mu_\star\right)\right) R - \lambda^\text{4} \mathcal{F}^{\text{eff}}_\lambda\left(\frac{\bar m \sigma_2}{\lambda^2}; \mu_\star\right) \right]  \\
+ \int d^{\text{4}}x \sqrt{g} \left[ \frac{1}{2M_{P}^{2}}\left( \mathcal{F}^{\text{eff}}_{\text{GB}, 1} \left(\frac{\sigma_1}{M_{P}} ; \mu_\star\right)+\mathcal{F}^{\text{eff}}_{\text{GB}, 2} \left(\frac{\sigma_2}{M_{P}} ; \mu_\star\right)\right) R_{\text{GB}} \right]  \\
+ \sum_{i=1}^2-Q_i C_i +\int d^{\text{4}}x \sqrt{g} \left[ -\frac{1}{2}  (\nabla\sigma_i )^2 -\frac{1}{2} (\mu_i \sigma_i+Q_i)^2 \right] \,,
\end{multline}
where
\begin{subequations}
\begin{align}
\mathcal{F}^{\text{eff}}_\lambda\left(X ; \mu_\star\right)&= \alpha^{\text{eff}}+\sum_{k\geq1} \beta^{\text{eff}}_k X^k\,, \\
 \mathcal{F}^{\text{eff}}_{R,i} \left(X ; \mu_\star\right)&= \sum_{k\geq1} c_k^{(i,\text{eff})}  X^k \,,\\
 \mathcal{F}^{\text{eff}}_{R_{\text{GB}},i}\left(X ; \mu_\star\right)&= \sum_{k\geq1} d_k^{(i,\text{eff})}  X^k  \,, 
\end{align}
\end{subequations}
  and we defined {\em effective} coupling constants as functions of the physically measurable quantities. Explicitly,    
\begin{subequations}
\begin{align}
\nonumber
\alpha^{\text{eff}}&= \alpha_r (\mu_\star)  - \frac{\hbar}{32\pi^2} \bigg[ \sum_i  \frac{\mu_i^4}{2\lambda^4} \left[\frac32- \ln \left(\frac{\mu_i^2}{\mu_\star^2}\right)\right]  + 2 \beta_{r,2}(\mu_\star) \left[1- \ln \left(\frac{\mu_2^2}{\mu_\star^2}\right)\right]  \frac{ \bar m^2 \mu_2^2}{\lambda^4} \\
& \hspace{2.8cm} - 2\beta_{r,2}^2(\mu_\star) \ln \left(\frac{\mu_2^2}{\mu_\star^2}\right)\frac{\bar m^4}{\lambda^4} \bigg] \\
\beta_j^{\text{eff}}&= \beta_{r,j}(\mu_\star) - \frac{\hbar}{32\pi^2} \bigg[  \left(1- \ln \left(\frac{\mu_2^2}{\mu_\star^2}\right)\right)   (j+2)(j+1) \beta_{r,j+2}  \frac{ \bar m^2 \mu_2^2}{\lambda^4}  \nonumber \label{rad_beta}\\
&-\frac{1}{2} \ln \left(\frac{\mu_2^2}{\mu_\star^2}\right) \frac{\bar m^4}{\lambda^4} \sum_{l=0}^{j} (l+1) (l+2 )(j-l+2)(j-l+1) \beta_{r,j-l+2}(\mu_\star)\beta_{r,l+2}(\mu_\star)   \bigg] \\
c^{(i,\text{eff})}_j &= c_{r,j}^{(i)}(\mu_\star) + \frac{\hbar}{32 \pi^2 } \delta_{i2}\ln \left(\frac{\mu_2^2}{\mu_\star^2}\right) \left(c^{(2)}_{r,2}(\mu_\star) + \frac16 \right)  2 (j+2)(j+1) \beta_{r,j+2}  \frac{\bar{m}^2}{M_P^2} \left(\frac{\bar{m} \, M_{P}}{ \lambda^2}\right)^j  \label{rad_c}\\
\d_j^{(i,\text{eff})}&= d^{(i)}_{r,j}(\mu_\star) - \frac{\hbar}{32\pi^2} \bigg[  \left(1- \ln \left(\frac{\mu_2^2}{\mu_\star^2}\right)\right)  \frac{1}{2} (j+2)(j+1) d_{r,j+2}^{(i)}  \frac{ \mu_i^2} {M_P^2}  \nonumber \label{rad_d}\\
&-\frac{1}{2} \delta_{i2}\ln \left(\frac{\mu_2^2}{\mu_\star^2}\right)  \sum_{l=0}^{j} (l+1) (l+2 )(j-l+2)(j-l+1) \d^{(2)}_{r,l+2}(\mu_\star)\beta_{r,j-l+2}(\mu_\star) \frac{\bar{m}^2}{M_P^2} \left(\frac{\bar{m} \, M_{P}}{ \lambda^2}\right)^{j-l}  \bigg] \\\nonumber
Z^{\text{eff}} &= Z_r(\mu_\star) - \frac{\hbar}{32\pi^2 } \bigg[ \sum_i 2\mu_i^2 \left(c_{ r,2}^{(i)}(\mu_\star)+\frac16\right) \left[ 1- \ln \left(\frac{\mu_i^2}{\mu_\star^2}\right) \right]\\\nonumber
&\hspace{2.8cm}- 4 \left( c_{r,2}^{(2)}(\mu_\star)+\frac16 \right)  \beta_{r,2}(\mu_\star) \bar m^2 \ln \left(\frac{\mu_2^2}{\mu_\star^2}\right) \bigg]\frac{1}{M_P^2} \label{Zrun} \\
u^{\text{eff}} &= u(\mu_\star) + \frac{\hbar}{32\pi^2}\sum_i \left[ \frac{1}{180} -\frac12 \left(c_{ r,2}^{(i)}(\mu_\star)+\frac{1}{6}\right)^2 \right]  \ln \left(\frac{\mu_i^2}{\mu_\star^2}\right) \\
v^{\text{eff}} &= v(\mu_\star) + \frac{\hbar}{32\pi^2}\left[-\frac{1}{60}\right] \sum_i  \ln \left(\frac{\mu_i^2}{\mu_\star^2}\right) 
\end{align}
\end{subequations}
Let us comment finally on several important issues:

\begin{itemize}

\item Radiative stability: We can now investigate the issue of radiative stability of the various couplings in the theory. Looking at the beta functions in the appendix, we easily notice that provided we take 
\begin{subequations}
\label{eq:scales}
\begin{align}
M_P \gtrsim \lambda \gtrsim \mu_1 \gtrsim \mu_2 \geq \bar{m} \;, \label{eq:scales_1}\\
\frac{\bar{m} M_P}{\lambda^2} \lesssim 1\label{eq:scales_2}
\end{align}
\end{subequations}
all beta functions remain small in the perturbative regime. In particular, all potentially dangerous mass ratios are naturally $\order(1)$. This remains true even if we consider the contribution of ordinary matter fields to the running of $\alpha$, provided $\lambda$ is as high as the cutoff of the effective theory. Moreover, from \eqref{rad_beta} and \eqref{rad_c} it can be inferred that the choice $\beta_{k \geq 5} =0$ and $c_{k\geq 3}^{(i)}=0$ is stable as advertised. Also note that condition \eqref{eq:scales_2} disappears if we truncate $\mathcal{F}_{R_{GB},2}$ at quadratic order (corresponding to $d^{(2)}_{j \geq 3} =0$) as follows from \eqref{rad_d}.

\item Effective action truncation: The expansion of the effective action we utilized originates from the asymptotic expansion of the heat kernel in Eq.~(\ref{DWexp}), which necessarily does not converge. One may nevertheless view the result as a {\em derivative} expansion, with the cutoff being the axion masses $\mu_i$. In order that the expansion is meaningful, we need ask how good it is to ignore $E_3(x)$, Eq.~(\ref{E3}), from our analysis of sequestering in the quantum-corrected theory. In curved space, however, one should be precise about the meaning of the derivative expansion because general covariance makes it difficult to define an expansion parameter. Nevertheless, we can define a {\em local} relational measure as follows
\begin{align}\label{curvature_constr}
\frac{\{\mathcal{R}^3, \nabla \nabla \mathcal{R}^2, \Box \Box R \}}{\mu_i^2} \ll \{\mathcal{R}^2 \} 
\end{align}
which dictates that in units of axion masses the curvature invariants forming $E_3(x)$ are subdominant to those in $E_2(x)$. Still this is not the full story, since the background values of the axions must be constrained as well to ensure the validity of the truncation. Explicitly
\begin{align}
 \frac{\mathcal{Q}_i }{\mu_i^2} \ll 1
\end{align}
which then translates into the following constraints
\begin{subequations}
\begin{align}\label{convergence}
\F_\lambda'' \left(\frac{\bar m \bar \sigma_2}{\lambda^2 }\right) \lesssim 1 \,,\\
\F_{O, i}''\left(\frac{\bar \sigma_i}{M_P}\right) \lesssim 1 \,.
\end{align}
\end{subequations}
where we have assumed that $\mathcal{R} \ll \mu_i^2$ consistent with \eqref{curvature_constr}.
\item In order to make contact with both sequestering models discussed in section \ref{sec:review}, we specify two subclasses of \eqref{action4} that are both stable under axion loops. To be precise, we recover the old sequestering proposal by setting $d_{j}^{(i)} =0$, and we obtain \textit{Omnia Sequestra} by demanding $c_j^{(i)}=0$, $d_{j\geq 3}^{(i)}=0$ and $\beta_{j\geq 2}=0$. Both models, in the absence of quantum gravity corrections, correspond to 1-loop exact, renormalisable theories.

\end{itemize}

\section{The robustness of sequestering under quantum corrections} \label{sec:robust}
We are now in a position to test the robustness of the sequestering mechanism after correcting the action by axion loops.  For both classes of model, the quantum corrected action can be written in the compact form
\begin{eqnarray} \label{actionq}
S[ \sigma_i,  g, \Psi ; Q_i) &=& \int d^{\text{4}}x \sqrt{g}   \left[ \frac{M_P^2}{2}  Z R+u R^2 + v R_{\mu\nu} R^{\mu\nu}  
\right]+S_m [g_{\mu\nu}, \Psi]\nonumber  \\
&+& \int d^{\text{4}}x \sqrt{g} \left[ \sum_{O=R,R_\text{GB}}  \frac{M_P^{4-[O]}}{2}\left\{ \mathcal{F}_{O, 1} \left(\frac{\sigma_1}{M_{P}} \right)+\mathcal{F}_{O, 2} \left(\frac{\sigma_2}{M_{P}} \right)\right\} O[g] - \lambda^\text{4} \mathcal{F}_\lambda\left(\frac{\bar m \sigma_2}{\lambda^2}\right) \right] \nonumber \\
&+& \sum_{i=1}^2-Q_i C_i +\int d^{\text{4}}x \sqrt{g} \left[ -\frac{1}{2}  (\nabla\sigma_i )^2 -\frac{1}{2} (\mu_i \sigma_i+Q_i)^2\right]\,,
\end{eqnarray}
where
\begin{subequations}
\label{Fs2}
\begin{align}
\mathcal{F}_\lambda(X)&= \alpha+\sum_{k\geq1}^{4} \beta_k X^k\,, \label{Fs2_lambda} \\
 \mathcal{F}_{R,i} (X)&= \sum_{k=1}^2 c_k^{(i)}  X^k \,,\label{Fs2_R}\\
 \mathcal{F}_{R_{\text{GB}},i} (X)&= \sum_{k\geq1} d_k^{(i)}  X^k  \label{Fs2_GB}\,,
\end{align}
\end{subequations}
and the couplings are understood to be the effective couplings, although we drop the superscript ``eff" for brevity.  We will now proceed in a way similar to section \ref{sec:review}.  The field equations  now yield 
\begin{eqnarray}
\sum_{O=R,R_\text{GB}}    \frac{M_P^{3-[O]}}{2} \mathcal{F}_{O,1}' \left(\frac{\sigma_1}{M_P}\right) O[g] &=& -\square \sigma_1+{\mu_1}\left(\mu_1\sigma_1+Q_1 \right)\label{sc1q}\\
\sum_{O=R,R_\text{GB}}    \frac{M_P^{3-[O]}}{2} \mathcal{F}_{O,2}' \left(\frac{\sigma_2}{M_P}\right) O[g] -\bar m \lambda^2  \mathcal{F}_\lambda'\left(\frac{\bar m \sigma_2}{\lambda^2 }\right) &=&  -\square\sigma_2+{\mu_2}\left(\mu_2\sigma_2+Q_2\right) \label{sc2q} 
\end{eqnarray}
and
\begin{multline}
 \mpl^2Z G_{\mu\nu}+\sum_{O=R,R_\text{GB}}   M_P^{4-[O]} E^O_{\mu\nu}(\mathcal{F}_{O,1}+\mathcal{F}_{O,2}) + H_{\mu\nu} 
=\\
 -\lambda^4 \mathcal{F}_\lambda g_{\mu\nu}+T_{\mu\nu}  +\sum_{i=1}^2 2 \nabla_\mu \sigma_i \nabla_\nu \sigma_i-g_{\mu\nu} \left[(\nabla\sigma_i )^2+\frac{1}{2} (\mu_i \sigma_i+Q_i)^2\right]  \label{gq}
 \end{multline}
 where $T_{\mu\nu}$ is the energy momentum tensors for matter and axions, defined in section \ref{sec:review}. Recall that we also defined a linear operator of the form
  \begin{eqnarray}
 E^O_{\mu\nu}(\mathcal{F})=\begin{cases} (-\nabla_{\mu} \nabla_{\nu} +g_{\mu\nu} \square +G_{\mu\nu} )\mathcal{F} & \text{for~} O[g]=R \\
 -4P_{\mu\alpha\nu\beta} \nabla^\alpha \nabla^\beta \mathcal{F}  &  \text{for~}  O[g]=R_{GB}  \end{cases}
  \end{eqnarray}
The quadratic curvature corrections contribute a term
\begin{multline}
{H_{\mu\nu}=2 u \, \left(2R R_{\mu\nu} - \frac12 g_{\mu\nu} R^2+2 g_{\mu\nu} \square R-2 \nabla_\mu\nabla_\nu R\right)}\\ {+2 v  \, \left( \nabla_\alpha \nabla_\beta R^{\alpha\beta} g_{\mu\nu} -2 \nabla_\alpha \nabla_{(\mu} R^\alpha_{\nu )}+\square R_{\mu\nu}+2 R^\alpha_\mu R_{\nu \alpha}-\frac12 R_{\alpha \beta} R^{\alpha\beta}g_{\mu\nu}\right)}
\end{multline}
Variation with respect to the global Lagrange multipliers yields the global constraints \eqref{C}, given in terms of the flux of the 3-forms through the boundary, $C_i=\int_{\partial \mathcal{M}} A_i$. It is convenient to adjust these  constraints  to  the form \eqref{Cmod} by  a redefinition of the 3-forms, $A_i \to (A_i-\star d \sigma_i)/{\mu_i}$, as described in section \ref{sec:review}.  Integrating the scalar equations of motion \eqref{sc1q} and \eqref{sc2q} over spacetime and using the adjusted form of the constraint \eqref{Cmod}, we obtain the following algebraic condition on the (weighted) spacetime average of the Ricci scalar,
\be \label{globconq}
\mathcal{A} \langle R \rangle_w^2+\mathcal{B} \langle R \rangle_w+\mathcal{C}=0
\ee
where 
\begin{eqnarray}
\mathcal{A}&=&\frac{1}{12 \mpl} \left \langle  \frac{K_{GB}}{w} \right \rangle_w \\
\mathcal{B} &=&\frac{1}{6\mpl} \left \langle  \frac{K_{GB}\delta R }{w} \right \rangle_w +\frac{\mpl}{2} \left \langle  \frac{K_{R}}{w} \right \rangle_w\\
\mathcal{C} &=& \frac{1}{2\mpl}\left \langle  \frac{K_{GB}\delta R }{w} \left(W_{\mu\nu\alpha\beta}^2  -2\left(R_{\mu\nu}-\frac14 R g_{\mu\nu}\right)^2 + {\frac{1}{6}}(\delta R)^2 \right) \right \rangle_w \nonumber \\
&&\qquad + \frac{\mpl}{2} \left \langle  \frac{K_{R}\delta R}{w} \right \rangle_w+\bar m \lambda^2\left\langle \frac{\F_\lambda'}{w} \right\rangle_w \frac{C_1}{C_2}
\end{eqnarray}
and we have introduced the shorthand $K_O \myeq\F_{O, 1}'-\frac{C_1}{C_2} \F_{O,2}'$ for $O=R, R_\text{GB}$. The choice of weighting factor, $w$, is not especially important, as long as it is built from radiatively stable quantities.  To make contact with section \ref{sec:review}, a natural choice would be 
$$
w=\begin{cases} K_R & \text{if $K_{GB} =0$}  \\K_{GB} & \text{if $K_{GB} \neq 0$} \end{cases}
$$
thereby eliminating some of the terms linear in $\delta R=R-\langle R\rangle_w$.  The constraint equation \eqref{globconq}  should still be enough to fix the large wavelength mode of the scalar curvature in a radiatively stable way, provided the coefficients $\cal A, B, C$ are themselves radiatively stable. Further, it is clear that it can be fixed to an arbitrarily  small value with a judicious, and radiatively stable,  choice of the flux ratio, $C_1/C_2$. The long wavelength mode of $\lambda^4 \mathcal{F}_\lambda+\sum_i \frac{1}{2} (\mu_i \sigma_i+Q_i)^2$ will again play the role of the cosmological constant counter term, forced by \eqref{globconq} to take on precisely the right value to cancel off radiative corrections to vacuum energy.  Let us see this explicitly by calculating the effective gravity equation. By taking traces and (weighted) spacetime averages of \eqref{gq}, we find that
   \begin{multline}
  M_\text{eff}^2 G_{\mu\nu}+\sum_{O=R, R_\text{GB}}M_P^{4-[O]} E^O_{\mu\nu}(\delta \mathcal{F}_{O,1}+\delta \mathcal{F}_{O,2})+H_{\mu\nu} =\\
T_{\mu\nu}-\frac14 \langle T\rangle_w g_{\mu\nu} -\delta \left[\lambda^4 \mathcal{F}_\lambda+ \sum_{i=1}^2 \frac{1}{2} (\mu_i \sigma_i+Q_i)^2\right] g_{\mu\nu} 
+\sum_{i=1}^2  \left[2 \nabla_\mu \sigma_i \nabla_\nu \sigma_i-g_{\mu\nu} (\nabla\sigma_i )^2  \right]- \Lambda_\text{global} g_{\mu\nu} \label{effgq} 
 \end{multline}
where we define $\delta \chi= \chi-\langle  \chi  \rangle_w$ as the {\it localised} fluctuation in a generic field $\chi$, in terms of the weighted averages with weight $w$. The effective Planck mass is  
 \be
M_\text{eff}^2 = \mpl^2(Z+ \langle \mathcal{F}_{R, 1}+ \mathcal{F}_{R, 2}\rangle_w) 
\ee
As before, the potential contains a localised part, corresponding to  $\delta \left[ \lambda^4 \mathcal{F}_\lambda +\ldots \right] $, and a global, infinite wavelength part, given by 
\begin{multline}
\Lambda_\text{global}=\Lambda_*-\frac14\left\{\sum_{O=R, R_\text{GB}} M_P^{4-[O]} \left[\langle  E^O(\delta \mathcal{F}_{O,1})+  E^O( \delta \mathcal{F}_{O,2}) \rangle_w\right]\right. \\ \left.
+2\sum_{i=1}^2 \left \langle( \nabla {\sigma_i} )^2 \right\rangle_w -2(3u-v)\langle \square R \rangle_w \right\} \label{Lglobq}
\end{multline}
where  $\Lambda_*=\frac14 M_\text{eff}^2  \langle R \rangle_w $ with the corresponding formula for   $\langle R \rangle_w$ now given by the solution to \eqref{globconq}. To examine the impact of radiative corrections to the vacuum energy most succinctly, it is convenient to go to the global limit  in which the spacetime geometry, the scalars and the matter source  are all maximally symmetric, as we did in section \ref{sec:review}. In other words, we have constant curvature, $R_{\mu\nu\alpha\beta}=\frac{\bar{R}}{12} (g_{\mu\alpha} g_{\nu\beta}- g_{\mu\beta} g_{\nu\alpha})$,  constant scalars, $\sigma_i=\bar \sigma_i$, and a vacuum energy source, $T_{\mu\nu}=-V_\text{vac} g_{\mu\nu}$. Since $H_{\mu\nu}$ vanishes on these configurations the effective gravity equation will once again yield an expression for the curvature given as $\bar R=\langle R \rangle_w$, with $\langle R \rangle_w$  constrained by the fluxes according to equation \eqref{globconq}. The only question is whether or not the solution to this equation is guaranteed to be radiatively stable. This will indeed be the case provided the coefficients $\cal A,  B, C$ are themselves radiatively stable. Now $\bar m$, $\lambda$ and $\mpl$  are fixed scales so they don't receive any corrections\footnote{Recall that the running of the gravitational coupling has been parametrised through  $Z$. This is radiatively stable against matter loops for sub-Planckian field theory cut-offs \cite{Demers} and  against axion loops as long as $\mu_i \lesssim  \mpl$, as can be seen from \eqref{Zrun}}.  The fluxes, $C_i$,  are set as infra-red boundary conditions, so the only possible source of ultra-violet sensitivity in \eqref{globconq} is through the spacetime averages of the combinations of (derivatives of) $\F_{O, i}$ and $\F_\lambda$. In the global limit, the spacetime averages just reduce to their global values. Now, as argued in the previous section, the couplings {$c_j^{(i)}$, $d_j^{(i)}$ and $\beta_j$}, corresponding to the coefficients of  the defining functions \eqref{Fs2}, are at most order one and  stable under axion loops {if conditions \eqref{eq:scales} are fulfilled. In particular $\lambda \gtrsim \mu_1 \gtrsim \mu_2$, which} we can also  take as an indication of the stability under matter loops, provided that the scale $\lambda$ lies at or above the field theory cut-off. 

What about the size and {sensitivity to changes in the vacuum energy} of the arguments $\frac{\sigma_1}{\mpl}$, $\frac{\sigma_2}{\mpl}$ and $\frac{\bar m \sigma_2}{\lambda^2}$? To study this it is convenient to consider the  original equations \eqref{sc1q} to \eqref{gq}  in the global limit, 
 \begin{eqnarray}
\mu_1 (\mu_1 \bar \sigma_1 +Q_1) &=&\frac{\mpl}{2} \F_{R, 1}' \left(\frac{ \bar \sigma_1}{\mpl }\right) \bar R+\frac{1}{12 \mpl} \F_{R_{\text{GB}},1}' \left(\frac{ \bar \sigma_1}{\mpl }\right) \bar R^2 \label{scglob1}\\
 \mu_2 (\mu_2 \bar \sigma_2 +Q_2)+\bar m \lambda^2  \mathcal{F}_\lambda'\left(\frac{\bar m \bar \sigma_2}{\lambda^2 }\right) &=&  \frac{\mpl}{2} \F_{R, 2}' \left(\frac{ \bar \sigma_2}{\mpl }\right) \bar R+\frac{1}{12 \mpl }\F_{R_{\text{GB}},2}' \left(\frac{ \bar \sigma_2}{\mpl }\right)\bar R^2\qquad \label{scglob2}\\
 M_\text{eff}^2 \bar R &=& 4 \left(\lambda^4 \mathcal{F}_\lambda +\sum_i\frac{1}{2} (\mu_i \bar \sigma_i+Q_i)^2+V_\text{vac}\right) \label{globg}
 \end{eqnarray}
Given  the smallness of the observed value of $\mpl^2 \bar R$ in terms of $V_\text{vac}$  we can infer good approximations for the $\bar \sigma_i$ by studying this equation in the limit of vanishing $\bar R$. {\it Assuming}  $\frac{\bar m \sigma_2}{\lambda^2} \lesssim 1$, one can straightforwardly show that
\begin{eqnarray}
\bar{\sigma}_1 &\approx& -Q_1/\mu_1 \\
 \bar{\sigma}_2 & \approx &\order(1) \left[1+\order(1) \frac{V_\text{vac}}{\lambda^4} \right] \frac{\lambda^2}{\bar m} \label{sig2sol}
\end{eqnarray}
Our approximation is self-consistent as long as  $\lambda$ lies at or above the field theory cut-off, ensuring that $V_\text{vac} \lesssim \lambda^4 $. We have also used the fact that $\bar m \ll \mu_2$ {[required for the convergence of the heat kernel expansion, cf.\ \eqref{convergence}]} and that the functions contain order one couplings in their definitions \eqref{Fs2}.  We now see how the axion field values have no UV sensitivity to leading order. Of course, the subleading corrections to $\bar \sigma_i$ do depend on the vacuum energy $V_\text{vac}$ but the point is that the sensitivity is weak thanks to the assumptions we have made on the parameters of the theory. An order one rescaling of the vacuum energy, $\Delta V_\text{vac} =\order(1) V_\text{vac}$, will only induces changes in the axion that go as {$\Delta \bar \sigma_2=\order(1) \bar \sigma_2 \Delta V_\text{vac} /\lambda^4 \lesssim \order(1) \bar \sigma_2$ and leave $\bar{\sigma}_1$ unchanged}.

We conclude that the quantum corrected theory \eqref{action} retains the ability to sequester radiative corrections to the vacuum energy arising from loops of matter.  This is a non-trivial result since the quantum corrections have introduced a new structure. As in section \ref{sec:review}, we have simplified our analysis by focussing on the global limit, in which we consider maximally symmetric configurations.  As explained then, the local fluctuations that we have ignored are irrelevant for considering the effects of vacuum energy since they cannot be sourced by it.  Their only contribution to the geometry on large scales is indirect, through their contribution to the spacetime averages entering in \eqref{Lglobq}. 

There is one last thing to check. In deriving the quantum corrected  action \eqref{actionq},  we made an assumption that  the second derivatives satisfied $\F_\lambda'' \left(\frac{\bar m \bar \sigma_2}{\lambda^2 }\right), ~\F_{O, i}''\left(\frac{\bar \sigma_i}{M_P}\right) \lesssim 1$, as required for a convergent heat kernel expansion. The condition on $\F_{R, i}''$ holds trivially given {its quadratic form},  with order one couplings, since $\F_{R, i}''\left(\frac{\bar \sigma_i}{M_P}\right) = 2 c^{(i)}_2 \sim \order(1)$. For $F_\lambda'' $ we take the form of \eqref{Fs2_lambda}, with order one couplings, and plug in the solution \eqref{sig2sol} for $\bar \sigma_2$ giving  
$
\F_\lambda'' \left(\frac{\bar m \bar \sigma_2}{\lambda^2 }\right) \sim \order(1)
$.
To guarantee $\F_{R_\text{GB}, i}'' \left(\frac{\sigma_2}{\mpl} \right) \lesssim \order(1)$, we generically require $\lambda^2 \lesssim \bar m \mpl$ {in accordance with \eqref{eq:scales_2}}. However, since  $\mpl, \lambda \gtrsim   \mu_1 \gtrsim \mu_2 \gg \bar m$ this condition is tricky to satisfy. For example, if we were to  take $\bar m $ to be at the GUT scale ($\sim 10^{16} $ GeV), consistent with high scale inflation, the condition becomes  $\lambda \lesssim 10^{17}$ GeV and so we {can only marginally} satisfy $\lambda \gtrsim \mu_i \gg \bar m$.\footnote{{Note that this condition is only a technical one needed to ensure the convergence of the heat kernel expansion. A priori, a situation where $\lambda \gtrsim \mu_i \gtrsim \bar m$ might still be radiatively stable.}}  The alternative, of course, is to assume that {$\F_{R_\text{GB}, i}$} is at most quadratic {or to sacrifice the inflation scenario by choosing $\bar{m}$ below the GUT scale}. In any event,  we  conclude that the assumptions made in deriving the effective action in the previous section can be made  self consistent.
 
 \section{Conclusions} \label{sec:conc}
 The extension of vacuum energy sequestering models \citeseq  to field theory models of monodromy \cite{monseq, irat} was an important step in the long term goal of realising the sequestering mechanism as an emergent phenomenon in a fundamental UV complete theory of Nature. However, this development also brought in additional moving parts, and it is important to ask whether or not they really are compatible with the original low energy models of sequestering, and more importantly, the vacuum energy cancellation mechanism itself.  In this paper we have answered both of these questions, having calculated the effect of loop corrections due to those additional moving parts, the axions.  The result is that the original low energy models of sequestering are incomplete because the axion loops necessarily generate new structures not seen in those original theories. However, as we have seen, for a new class of Lagrangians, closed under renormalisation up to a particular order in power counting, the cancellation of vacuum energy remains intact. As a result, the sequestering {\it mechanism} does seem  to be robust against quantum corrections from axion loops. 
 
 The details of our analysis relied on a number of reasonable assumptions regarding the parameters in the theory.  For the quantum corrected actions of the form \eqref{actionq}, we ought to assume the following hierarchy of scales
 \be
\mpl , \lambda \gtrsim \M,   \mu_1 \gtrsim \mu_2 \gg \bar m, k, \qquad \frac{\bar{m} M_P}{\lambda^2} \lesssim 1
\ee
where $\M$ is the field theory cut-off and $k$ is the momentum scale. We also assume that the dimensionless couplings in the definitions \eqref{Fs2} of $\F_{O, i }$ and $\F_\lambda$ are at most order one. Some of these conditions are only required to ensure the convergence of the heat kernel expansion (in particular, $\mu_i \gg \bar m, k$), while the rest play some role in ensuring the robustness of the sequestering mechanism. For example,  the scale $\lambda$ is always chosen to lie at or above the cut-off $\M$ in order to avoid large contributions in the effective gravity equation that go as $V_\text{vac}/\lambda^4$ with $V_\text{vac}  \sim \M^4$. Although the choice of scales is important, our analysis in section \ref{sec:robust} does seem to suggest that the precise {\it structure} of the monodromy deformation is  less important from the point of view of a successful sequestering mechanism. 

{Indeed, as long as the axion dependence contains suitable scalings, we anticipate the sequestering mechanism to survive  a variety of deformations to the specific examples we have studied here. These may include (i) couplings between the axion sectors; (ii) higher powers of the axion fields ($\beta_{j\geq 5}\neq 0$ and $c_{j \geq 2} \neq 0$); and even  (iii) a completely new type of coupling between $\sigma_i$ and higher curvature invariants (like $\sigma_i R^2$). 
Although our analysis was not quite as general as this, it still provided a non-trivial challenge to  the classical mechanism having included a consistent subset of loop-induced operators. These terms, which have not been discussed before in a classical context, necessarily arise from integrating out axion loops and we have shown that they leave sequestering intact. }

When, then, do we expect the sequestering mechanism to fail? The answer to this is suggested by the arguments in favour of {\it Omnia Sequestra} \cite{KP4,omnia} over the original proposals \cite{KP1, KP2, KPSZ}. Problems occur when the form of the couplings themselves become too UV sensitive.  For example, in \cite{KP4}, it was argued that loop corrections involving virtual gravitons alongside virtual matter fields would introduce UV sensitivity in the form of  $\F_\lambda$. This introduced UV sensitivity into the global geometric constraint so that the residual cosmological constant present in the effective gravity equations itself became UV sensitive. In this paper, this problem never arose because we only considered pure axion loops, and there was no direct coupling between the axions and the matter fields. Of course, virtual gravitons can mediate such an interaction but they were not considered. A complete analysis including the impact of graviton loops will certainly reduce the space of monodromy models for which sequestering remains robust.  This will be left for future work, but for now we speculate that possibly only the {\it Omnia Sequestra}  model with a linear axion-Gauss-Bonnet coupling will survive, thanks to the exact shift symmetry such a theory would possess.  If such a symmetry is indeed crucial, it would then be important to ask how it can emerge from a consistent quantum theory of gravity.
 

\begin{acknowledgments}
{We would like to thank Peter Millington and David Stefanyszyn for many helpful discussions.} The work of B.K.E.\ is supported in part by the Science and Technology Facilities Council (grant number ST/P000819/1). A.P.\ and S. N.\ are  funded by a Leverhulme Trust  Research Project Grant. A. P. is also funded by an STFC Consolidated Grant. F.N.\ is supported by Villum Fonden Grant 13384. CP3-Origins is partially funded by the Danish National Research Foundation, grant number DNRF90. 
\end{acknowledgments}

 \appendix

 \section{Integrating out 3-forms and 4-forms} \label{app:forms}
 In deriving the action \eqref{action2}, we introduced the Lagrange multiplier fields $Q_i(x)$ and promoted the four-form field strengths to fields in their own right. The four-forms, only entering algabraically, and at quadratic order, were then integrated out, resulting in a mass term for $\sigma_i$. At the level of the path integral this gave rise to an irrelevant overall factor
\be
\mathcal{N}_{\tilde{F}_i} = \int \mathcal{D} \tilde{F}_i \, \exp \left[- i \int d^4x\, \sqrt{-g}\,\frac{1}{2} \, \frac{1}{4!} \, \tilde{F}_i^2 \right]\,    
\ee
where
\be
\tilde{F}^i_{\alpha\beta\gamma\delta}= F^i_{\alpha\beta\gamma\delta} + \frac{1}{\sqrt{-g}} \, \epsilon_{\alpha\beta\gamma\delta}  \left( \mu_i \, \sigma_i + Q_i \right)\,.
\ee
To arrive at the global form of the action given in \eqref{action3}, we must also integrate out the three-form fields, $A^i$. Focussing on the part of the relevant part of the action \eqref{action2}, we  have
\begin{align}\label{eq:S_3_form}
  \mathcal{S}_{A_i} &= -\frac{1}{3!} \int_{\mathcal{M}} d^4 x \, Q_i(x) \, \epsilon^{\alpha\beta\gamma\delta} \partial_{\alpha} A^i_{\beta\gamma\delta} \nonumber \\
                    &= - \frac{1}{3!}  \int_{\partial\mathcal{M}} d^3 \tilde{x}  \, Q_i(\tilde{x})\, n_{\alpha} \epsilon^{\alpha\beta\gamma\delta} \tilde{c}^i_{\beta \gamma \delta} +\frac{1}{3!}  \int d^4 x \partial_\alpha Q_i(x) \epsilon^{\alpha\beta\gamma\delta} A^i _{\beta\gamma\delta}
\end{align}
where $\tilde{c}^i_{\beta\gamma\delta} = A^i_{\beta\gamma\delta}|_{\partial \mathcal{M}}$ corresponding to Dirichlet boundary conditions for the 3-form fields\footnote{Note that this is consistent with $\delta A_i|_{\partial \mathcal{M}}=0$ when varying the action. Moreover, $\partial M$ can be understood as a regulator brane at a sufficiently distant point in space.}. The 3-form field can be integrated out which yields a functional delta function, explicitly
\be
 \int \mathcal{D} A_i \, \exp \left[ -  \frac{i}{3!} \int d^4 x \partial_\alpha Q_i(x) \epsilon^{\alpha\beta\gamma\delta} A^i_{\beta\gamma\delta} \right] = \delta \left(\partial_{\alpha} Q_i(x) \epsilon^{\alpha\beta\gamma\delta} / 3! \right) 
\ee
where the last expression is somewhat formal. We can give it a precise meaning by discretising  the $Q_i(x)$ path integration in the next step. For simplicity we focus on the one dimensional case, which can be generalised to four dimensions. Dropping the index $i$ for brevity,  note that the path integral measure is defined via
\begin{align}
\mathcal{D} Q = \lim_{N \to \infty}  \int \, dQ_0 dQ_1 \ldots dQ_N  
\end{align}
where each integration runs from $-\infty$ to $+ \infty$ and $Q_a\myeq Q(x_a)$ is evaluated on a discrete grid with spacing  $\Delta x = \left( x_{N} - x_{0} \right)/N$. For the functional delta we get
\begin{align}
  \delta\left(\frac{d Q}{d x}\right) \to \prod_{a=0}^{N-1} \delta\left( \frac{Q_{a+1}-Q_a}{\Delta x } \right)\,.
\end{align}
Putting both expressions together we find that
\begin{align}
  \int \mathcal{D}Q \delta\left( \frac{d Q}{dx} \right)
  &= \lim_{N\to \infty} \int dQ_0 dQ_1 \ldots dQ_N \prod_{a=0}^{N-1} \delta\left( \frac{Q_{a+1}-Q_a}{\Delta x} \right) \nonumber \\
  &= \left[\lim_{N \to \infty} \left(\Delta x\right)^N  \right] \, \int d Q \, .
\end{align}
The functional integration therefore collapses to an ordinary integration over $Q \equiv Q_0 = Q_1 = \ldots = Q_N $ corresponding to a constant $Q(x)=Q$. The factor in squared brackets does not depend on the fields and therefore is an irrelevant constant which drops out if we normalise our amplitudes.

We have thus fully integrated out the 3-form fields from the original action \eqref{action3}. Going back to \eqref{eq:S_3_form}, we see that only the boundary term keeps a memory of their original presence due to their non-vanishing boundary values $\tilde{c}^i_{\beta\gamma\delta}$. Using the constancy of $Q_i(x)$, we can write
\be
  \mathcal{S}_{A_i} = -Q_i C_i \, ,
\ee
where we defined $C_i=\frac{1}{3!}\int_{\partial \mathcal{M}} d^3\tilde{x} \,  n_{\alpha} \,  \epsilon^{\alpha\beta\gamma\delta} \tilde{c}^i_{\beta \gamma \delta}  $. Putting everything together, the final form of the sequestering action reads as \eqref{action3}, with $Q_i$ constant. These fields play the role of global variables reminiscent of the very first sequestering model \cite{KP1, KP2}. Note that at the level of path integral we retain the ordinary integration,  over these global variables, $dQ_i$, as discussed  at the beginning of section \ref{sec:effaction}.

\section{Counter terms} \label{CTls1}
As stated in the text, we carry out our renormalization in the $\overline{\text{MS}}$ renormalisation scheme. As usual, we introduce renormalization constants as follows
\begin{align}
\alpha = \alpha_r(\mu) + \delta_{\alpha},
\end{align}
and so on. We denote renormalised couplings with a subscript $r$, and they all run with the renormalization scale $\mu$. The bare action is given in Eq. \eqref{action4}. We find
\begin{subequations}
\label{deltas}
\begin{align}
\delta_\alpha&=\frac{\hbar}{32\pi^2 \bar{\epsilon}} \bigg[ \sum_i  \frac{\mu_i^4}{2\lambda^4}   + 2 \beta_{r,2}(\mu_\star)   \frac{ \bar m^2 \mu_2^2}{\lambda^4} + 2\beta_{r,2}^2(\mu_\star) \frac{\bar m^4}{\lambda^4} \bigg] \\
\delta_{\beta_j}&= \frac{\hbar}{32\pi^2 \bar{\epsilon}} \bigg[   (j+2)(j+1) \beta_{r,j+2}  \frac{ \bar m^2 \mu_2^2}{\lambda^4}  \nonumber\\
&\quad + \frac{1}{2} \frac{\bar m^4}{\lambda^4} \sum_{l=0}^{j} (l+1) (l+2 )(j-l+2)(j-l+1) \beta_{r,j-l+2}(\mu_\star)\beta_{r,l+2}(\mu_\star)   \bigg] \\
\delta_{c^{(i)}_j} &=  \frac{\hbar}{32 \pi^2 \bar{\epsilon}}\left(c^{(2)}_{r,2}(\mu_\star) + \frac16 \right)  2 (j+2)(j+1) \, \beta_{r,j+2} \,  \frac{\bar{m}^2}{M_P^2} \left(\frac{\bar{m} \, M_{P}}{ \lambda^2}\right)^j \delta_{i2} \\
\delta_{d_j^{(i)}}&=  \frac{\hbar}{32\pi^2\bar{\epsilon}} \bigg[    \frac{1}{2} (j+2)(j+1) d_{r,j+2}^{(i)}  \frac{ \mu_i^2} {M_P^2}  \nonumber \\
&+\frac{1}{2} \delta_{i2}  \sum_{l=0}^{j} (l+1) (l+2 )(j-l+2)(j-l+1) \d^{(2)}_{r,l+2}(\mu_\star)\beta_{r,j-l+2}(\mu_\star) \frac{\bar{m}^2}{M_P^2} \left(\frac{\bar{m} \, M_{P}}{ \lambda^2}\right)^{j-l}  \bigg] \\
\delta_Z &=  \frac{\hbar}{32\pi^2\bar{\epsilon} } \bigg[ \sum_i 2\mu_i^2 \left(c_{ r,2}^{(i)}(\mu_\star)+\frac16\right) + 4 \left( c_{r,2}^{(2)}(\mu_\star)+\frac16 \right)  \beta_{r,2}(\mu_\star) \bar m^2  \bigg]\frac{1}{M_P^2}  \\
\delta_u &=  \frac{\hbar}{32\pi^2\bar{\epsilon}}\sum_i \left[ \frac{1}{180} -\frac12 \left(c_{ r,2}^{(i)}(\mu_\star)+\frac{1}{6}\right)^2 \right]  \\
\delta_v &=  -\frac{\hbar}{32\pi^2\bar{\epsilon}}\left[\frac{1}{30}\right]   
\end{align}
\end{subequations}

We are now able to determine the beta functions of the various couplings from the coefficient of the pole in the above equations,
\begin{align}
\beta(X) = \bar{\epsilon} \, \delta_X \,,
\end{align}
 with $\delta_X$ as defined in \eqref{deltas} and
\begin{align}
   \beta(x) := \frac{dx}{d\log \mu^2} \ \ .
\end{align}

\end{document}